\documentclass[aps,prb,twocolumn,superscriptaddress,showpacs]{revtex4}

\usepackage{graphicx}  % needed for figures
\usepackage{dcolumn}   % needed for some tables
\usepackage{bm}        % for math
\usepackage{amssymb}   % for math
\usepackage{amsmath}

\begin{document}

%\preprint{}

%DIFFMARK

\title{\textit{Ab initio} prediction of stable nanotwin double layers and 4O structure in Ni$_{2}$MnGa}


\author{Martin Zelen\'{y}}
\email[]{zeleny@fme.vutbr.cz}
\affiliation{Institute of Materials Science and Engineering, NETME Centre, 
Faculty of Mechanical Engineering, Brno University of Technology, Technick\'{a} 
2896/2, CZ-616 69 Brno, Czech Republic}
\affiliation{Central European Institute of Technology, CEITEC MU, Masaryk 
University, Kamenice 753/5,CZ-625 00 Brno, Czech Republic}
\affiliation{Institute of Physics, Academy of Sciences of the Czech Republic, 
Na 
Slovance 1999/2, CZ-182 21 Prague, Czech Republic}

\author{Ladislav Straka}
\affiliation{Institute of Physics, Academy of Sciences of the Czech Republic, 
Na 
Slovance 1999/2, CZ-182 21 Prague, Czech Republic}

\author{Alexei Sozinov}
\affiliation{Material Physics Laboratory, Lappeenranta University of 
Technology, 
Laitaatsillantie 3, FI-57170 Savonlinna, Finland}

\author{Oleg Heczko}
\affiliation{Institute of Physics, Academy of Sciences of the Czech Republic, 
Na 
Slovance 1999/2, CZ-182 21 Prague, Czech Republic}

%\thanks{}


\date{\today}

\begin{abstract}
% insert abstract here.
The \textit{ab initio} electronic structure calculations of the Ni$_{2}$MnGa alloy indicate that the orthorhombic 4O structure exhibits the lowest energy compared to all known martensitic structures. The 4O structure is formed by
nanotwin double layers, i.e., oppositely oriented nanotwins consisting of two (101) lattice planes of nonmodulated martensitic structure. It exhibits the lowest occupation of density of states at the Fermi level. The total energy
1.98 meV/atom below the energy of nonmodulated martensite is achieved within structural relaxation by shifting Mn and Ga atoms at the nanotwin boundaries. The same atomic shift can also be found in other martensitic
nanotwinned or modulated structures such as 10M and 14M, which indicates the importance of the nanotwin double layer for the stability of these structures. Our discovery shows that the nanotwinning or modulation is a
natural property of low-temperature martensitic phases in Ni-Mn-Ga alloys.
\end{abstract}

%\keywords{}

\maketitle

Among Heusler alloys, the Ni$_{2}$MnGa is one of the few, which exhibit the 
unique properties resulting in a giant magnetic field-induced strain (MFIS) in 
their
low-temperature martensitic phase \cite{Ullakko96}. The MFIS can reach 
up to 12 \% in case of Co- and Cu-doped Ni$_{2}$MnGa 
\cite{Sozinov13}. 
Important factors enabling the MFIS are large magneto-crystalline anisotropy 
and 
extraordinarily high mobility of martensite twin boundaries \cite{Gruner08, Soderberg06, 
Heczko09, Acet11} that primarily depends on martensite structure. While at 
elevated temperature there is a single austenitic phase of Ni$_{2}$MnGa with 
cubic L2$_{1}$ structure, several low-temperature martensitic phases have been 
observed depending on composition, temperature and applied stress 
\cite{Martynov92, Lanska04, Cakir13}. Martensitic phases with orthorhombic or 
monoclinic structures exhibit modulation of (110) planes in $[1\overline{1}0]$ 
direction with the periodicity of ten or fourteen lattice planes (10M or 14M). 
The third martensitic phase with nonmodulated (NM) tetragonally distorted 
L2$_{1}$ structure is typical for larger deviation from stoichiometry. In 
addition, above the martensitic transformation temperature there is a cubic 
premartensite with the periodicity of six lattice planes (6M) in alloys with 
small deviation from stoichiometry \cite{Opeil08}.

Low-temperature instability of L2$_{1}$ austenite can be explained by a strong 
Fermi surface nesting responsible for the phonon softening in [110] direction, 
which explains structural modulation of 6M premartensite \cite{Zayak03, 
Entel06, 
Bungaro03, Zayak06}. In addition there is also a large Ni-$e_{g}$ peak 
corresponding to antibonding states right below the Fermi level, $E_{f}$, in the 
minority density of states (DOS) channel. Due to the band Jahn-Teller effect 
the 
L2$_{1}$ cubic structure lowers its symmetry, which pushes the peak above 
$E_{f}$ resulting in lower energy \cite{Fuji89, Brown99, Kart08}. Nonetheless, 
the precise causes of the variety of observed low temperature martensitic 
phases 
and mechanism of intermartensitic transformations between them have not been 
fully explained yet.

\textit{Ab initio} calculations have been employed to understand modulations in 
Ni$_{2}$MnGa. Early works used description of 10M and 14M martensite with the 
help of pseudo-tetragonal structures with harmonic modulation, but this 
approximation resulted in total energy approximately in the middle between 
energies of the austenite and NM structure \cite{Zayakjpcm, Zayak04, Luo12, 
Zayak05}. This is in conflict with apparent stable modulated structure observed 
for alloys close to stoichiometry at low temperatures \cite{Straka13, 
Heczko13}. 

\begin{figure*}
\includegraphics[width=150mm]{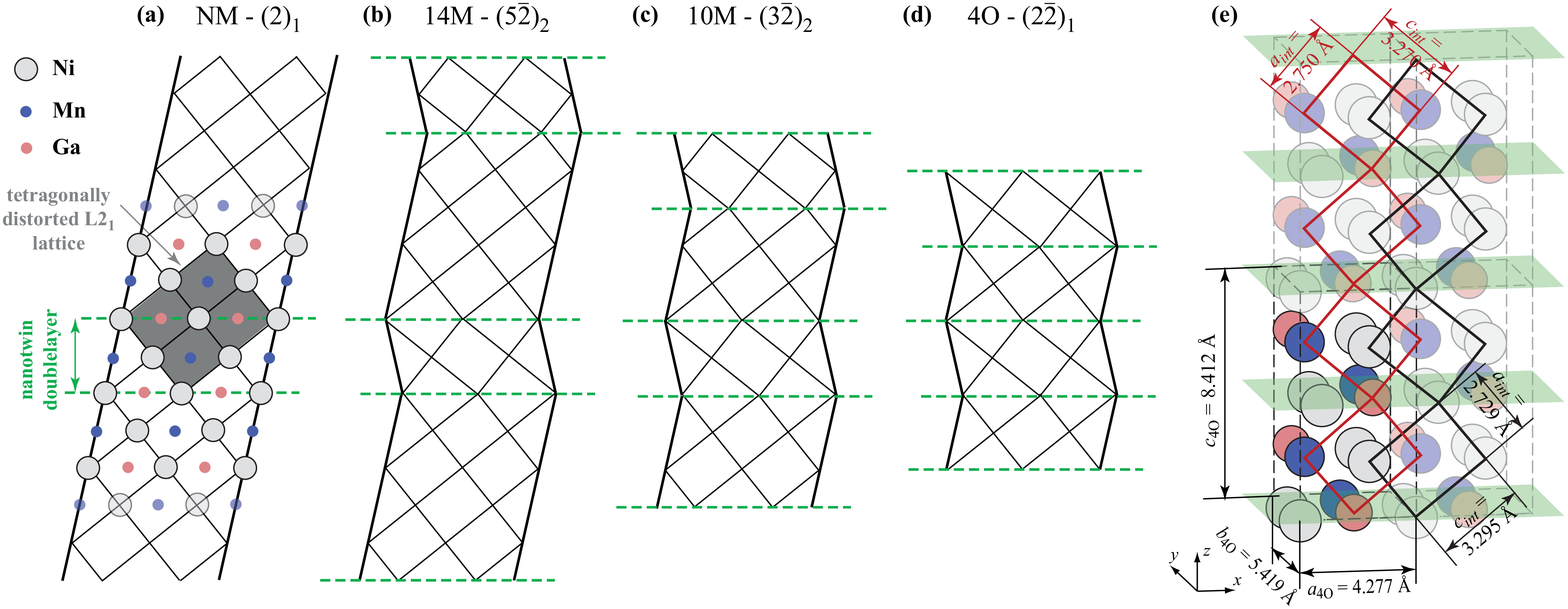}
\caption{\label{fig:fig1}(Color online) Schematic view of (a) NM, (b) 14M, (c) 
10M, and (d) studied 4O structure. The thin black lines mark tetragonal 
building blocks. One of the NM cells with tetragonally distorted L2$_{1}$ 
structure, consisting of eight tetragonal building blocks (only four visible), 
is exemplarily marked in gray. The green dashed lines mark doublelayer nanotwin 
boundaries constituted from (101) lattice planes of NM structure. (e) Detail 
description of calculated 4O structure ($2 \times 1 \times 2$ supercell 
displayed with highlighted unit cell) with lattice parameters $a_{\text{4O}}$, 
$b_{\text{4O}}$ and $c_{\text{4O}}$, and internal dimensions $c_{int}$ and 
$a_{int}$ of its tetragonal building blocks in Ni sublattices (black 
rectangles) 
and Mn-Ga sublattices (red rectangles). Green planes indicate nanotwin 
boundaries in (101) planes of NM structure. }
\end{figure*}

Later, the modulated martensites have been described by monoclinic
structures with an alternating sequence of nanotwins constituted from
(101) lattice planes of NM structure shown in
Fig. \ref{fig:fig1}(a). The 14M (denoted as $(5\overline{2})_{2}$,
Fig.  \ref{fig:fig1}(b)) comprises of five lattice planes in one
orientation and two planes in the other orientation. This is repeated
twice to fulfill the atomic ordering \cite{Pons00,Kaufmann10, Niemann16}. Similarly the 10M structure can be considered as
alternating nanotwins of width three and two lattice planes (denoted
as $(3\overline{2})_{2}$, Fig.  1(c)) \cite{Kaufmann11}. The nanotwin
description is based on the theory of adaptive martensite
\cite{Khachaturyan}, where modulated structures are derived from the
accommodation of the geometrical mismatch at the interface between
austenite and martensite \cite{Kaufmann10}. In contrast with
harmonic modulation the fully relaxed $(3\overline{2} )_{2}$ and $(5\overline{2}
)_{2}$ structures with zig-zag modulation give \textit{ab initio}
total energies very close to the total energy of NM structure within
0.7 meV/atom interval.

Yet, the succession of structures with decreasing total energy at 0 K depends on 
the calculation settings. The successions of 14M, 10M and NM structures 
\cite{Xu12}, 10M, 14M and NM structures \cite{Niemann12} and even 10M, NM and 
14M structures \cite{Dutta16} have been reported in contrast to experiment 
where 
only 10M structure has been observed for stoichiometric single crystal 
\cite{Brown99}. For off-stoichiometry composition the (10M $\rightarrow$) 14M 
$\rightarrow$ NM sequences of intermartensitic transformations occur with 
decreasing temperature \cite{Segui05, Straka02}. Recent \textit{ab initio} 
calculations of the Ni$_{2}$MnGa phase diagram which include lattice and 
vibrational degrees of freedom propose that the stability of martensite and 
intermartensitic transformation can be understood solely using thermodynamic 
concepts \cite{Dutta16}. However, such predictions for low-temperature part of 
phase diagram are still uncertain due to the small structural energy difference 
between martensitic phases.   

The $(3\overline{2})_{2}$ and $(5\overline{2})_{2}$ structures can be
conceptually transformed into the NM phase by reorientation of
nanotwins constituted from two basal planes -- nanotwin doublelayer
\cite{Pond12} as shown in Fig. \ref{fig:fig1}. The transformation can
be viewed as application of local shear in (101) plane along the
[10$\overline{1}$] direction \cite{Zeleny15}. The doublelayer nanotwin
reorientation was indicated in recent experimental in-situ study of
stress-induced 14M $\rightarrow$ NM transformation using high
resolution TEM \cite{Ge15}.  Transformation between
$(3\overline{2})_{2}$ and $(5\overline{2})_{2}$ structures can be
described similarly but mechanism is more complex, because it
includes cooperative reorientation of several nanotwin doublelayers.

The shear-based transformation concept does not include motion of twining 
dislocations observed in experiment, yet it shows the relevance of nanotwin 
doublelayer for the intermartensitic transformation and the stability of 
modulated structure. Gruner \cite{Grunerhab} conjectured that the structure 
based 
only on oppositely oriented double layers, denoted as 4O \cite{Niemann16}, might be even lower 
in 
energy by a few meV/f.u than other martensitic structures. Moreover the TEM 
reveals a very frequent stacking error of modulation in 14M structure, e. g. 
the 
sequence of $(5\overline{2}2\overline{2}5)$ \cite{Ge15} containing three 
nanotwin doublelayers, which indicates the stability of doublelayer 
multistacks. 
The 4O structure  has been suggested for Ni$_{50}$Mn$_{37}$Sb$_{13}$, 
Ni$_{50}$Mn$_{37.5}$Sn$_{12.5}$ and Ni$_{50}$Mn$_{35}$In$_{15}$ to explain 
experimental diffraction patterns \cite{Sutou04} but was never reported in Ni-Mn-Ga.

Thus, both theoretical concept and experiments indicate the importance
of nanotwin doublelayers. This paper presents \textit{ab initio}
study of the 4O structure formed by nanotwin doublelayers, newly
suggested in Ni$_{2}$MnGa alloy, and the comparison of its stability
with respect to austenite and known martensitic structures.

Presented \textit{ab initio} calculations were performed using the Vienna 
Ab-initio Simulation Package (VASP) \cite{vasp1, vasp2} in which the 
electron-ion interaction was described by projector-augmented wave potentials 
\cite{Blochl, paw}. The electronic orbitals were expanded in terms of plane 
waves with a maximum kinetic energy of 600 eV. We used the gradient-corrected 
exchange-correlation functional proposed by Perdew, Burke, and Ernzerhof 
\cite{pbe}. The Brillouin zone (BZ) was sampled using a $\Gamma$-point 
centered mesh with the smallest allowed spacing between \textit{k}-points equal 
to 0.1 \AA$^{-1}$ in each direction of the reciprocal lattice vectors. This 
setting ensured constant \textit{k}-points density in all our calculations. The 
integration over the BZ used the Methfessel-Paxton smearing method 
\cite{mpsmear} with 0.02 eV smearing width. Settings for \textit{k}-point 
density and smearing width were obtained with the help of adaptive smearing 
method \cite{ags}. The total energy was calculated with a high precision by 
convergence to 10$^{-7}$ eV per computational cell. Relaxation of the atomic positions and structural 
parameters were performed by the quasi-Newton algorithm, using the exact 
Hellmann-Feynman forces, and was considered to be converged after all forces 
dropped below 1 meV/\AA$^{-1}$. 

The calculated total energies for different structures are shown in Fig. 2(e). 
The differences between the austenite and martensitic structures are 4.48 
meV/atom for 10M martensite described by fully relaxed $(3\overline{2})_{2}$ structure, 4.54 
meV/atom for NM martensite, and 4.92 meV/atom for 14M martensite described by fully relaxed 
$(5\overline{2})_{2}$ structure. These results are in good agreement with 
the previous investigations \cite{Dutta16} although in some studies the NM 
martensite exhibits slightly lower energy than 14M \cite{Xu12, Niemann12}. In 
addition to the previously reported martensitic phases, we have found the 4O 
martensitic structure as displayed in Fig. 1(e), which exhibits significantly 
lower total energy about 6.52 meV/atom below the total energy of the austenite. 

\begin{figure*}
\includegraphics[width=110mm]{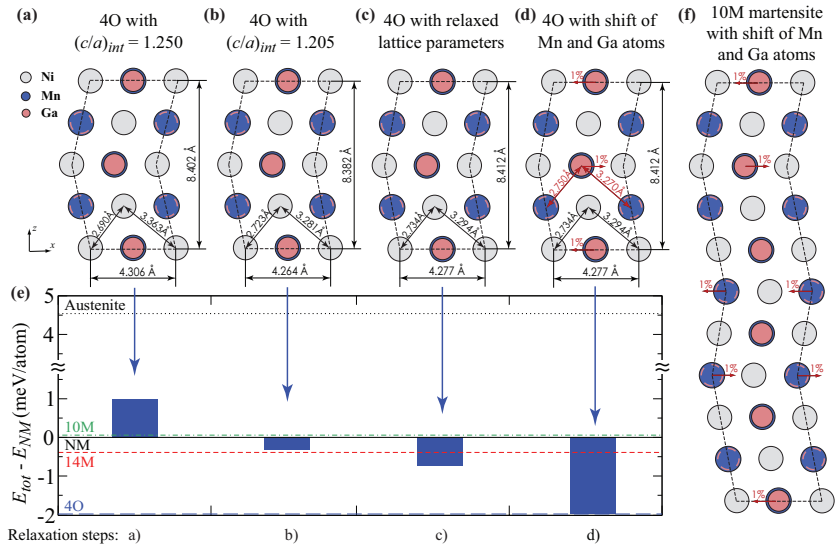}
\caption{
\label{fig:fig2}
(Color online) (a-d) Changes in the 4O
  structure during the relaxation process and (e) corresponding total
  energies for each structure with respect to the total energy of NM
  martensite. Total energies of fully relaxed 4O (blue dashed line),
  14M (red dashed line), 10M (green dash-and-dotted line) and
  austenite (black dotted line) are marked. Dashed circles mark Ga
  atoms in the atomic layer underneath the figure plane.  The small
  red arrows indicate the shift of Mn and Ga atoms in the nanotwin
  boundaries responsible for the significant total energy decrease. (f)
  Fully relaxed 10M structure with the shifts of Mn and Ga atoms marked.
}
\end{figure*}

The 4O structure in Fig. 1(e) constitutes from alternating doublelayer
nanotwins and thus can be denoted as $(2\overline{2})_{1}$. The
subscript indicates that the sequence is repeated only once. In
comparison to the ordinary NM structure with $(c/a)_{\text{NM}}$ =
1.250 determined by \textit{ab initio} calculations \cite{Zayak03},
the energy of 4O is 1.98 meV/atom lower (see
Fig. \ref{fig:fig2}(e)). Thus, our calculation indicates that neither
previously described modulated nor NM structures are the true ground
state of stoichiometric Ni$_{2}$MnGa alloy at 0 K.

The primitive cell of the 4O structure contains 16 atoms and exhibits
simple orthorhombic symmetry with $(c/a)_{\text{4O}} = 1.97$,
$(b/a)_{\text{4O}} = 1.27$ and lattice constant $a$ = 4.279
\AA.  These 16 atoms form two nanotwin doublelayers, which
are constituted from tetragonal building blocks with internal ratio
$(c/a)_{int}\approx 1.20$. This ratio is different from NM,
10M or 14M structures
\cite{Niemann12} with $(c/a)_{int}=1.25$. Creation of alternating
nanotwins doublelayers from $(c/a)_{int}=1.25$ blocks, which was
starting atomic configuration before relaxation (see Fig.
\ref{fig:fig2}(a)), leads to total energy 1.00 meV/atom above the NM
structure (see Fig.  \ref{fig:fig2}(e)). The lattice parameters of
this ideal structure were obtained from geometrical considerations \cite{Niemann16}
using the equations:

\begin{equation}
a_{\text{4O}} = a_{int}\sqrt{1+\left(\tfrac{c}{a}\right)_{int}^{2}}, 
b_{\text{4O}} = 2a_{int}, c_{\text{4O}} = 
\frac{4a_{int}^{2}\left(\tfrac{c}{a}\right)_{int}}{a_{\text{4O}}},
\end{equation}

where $(c/a)_{int} = (c/a)_{\text{NM}}$ and $a_{int} =
a_{\text{NM}}/2$. Atomic volume was kept constant and equal to the
equilibrium NM structure volume.

The ideal 4O structure is not stable and thus will relax to
  the above described ground state. This was achieved by full
  unconstrained relaxation of lattice parameters and atomic positions
  based on Hellmann-Feynman forces. However, for the purpose of
  pointing out the most important lattice changes responsible for
  stabilization of structure, we intentionally divided the relaxation
  process into several steps. During each step, only selected
  parameters or positions were allowed to change while remaining ones
  were artificially constrained. Both unconstrained and constrained
  approaches resulted in the same equilibrium structure.

In the first relaxation step,
 decreasing $(c/a)_{int}$ of the tetragonal
blocks to value 1.205 at the constant cell volume
(Fig. \ref{fig:fig2}(b)) decreased the total energy to the level 0.32
meV/atom below the NM structure, but still above the 14M
structure. The relaxation of lattice parameters accompanied by modest
volume expansion (Fig. \ref{fig:fig2}(c)) further decreased the total
energy to the level 0.72 meV/atom below the NM structure.  

The essential decrease of the total energy by additional 1.26 meV/atom
was achieved by distortion in Mn-Ga sublattice
(Fig. \ref{fig:fig2}(d)). This distortion arose from the shifts of Mn
and Ga atoms by about 1 \% of atomic coordinates along the $x$
direction. The shift occurred only for atoms at nanotwin boundary (red
arrows in Fig. \ref{fig:fig2}(d)) and its direction alternated at each
nanotwin boundary. Consequently the internal $(c/a)_{int}$ ratio of
tetragonal building blocks of Mn-Ga sublatttice decreased from 1.20 to
1.19 (red rectangles in Fig. \ref{fig:fig1}(e)).

Finally, the fully relaxed 4O structure was achieved by alternating
shift in Ni sublattice but not only along $x$ direction, but also
along $y$ direction and with magnitude less than 0.1 \% of atomic
coordinates.  Although this final relaxation step had negligible
effect on the total energy, it resulted in the increase of the
internal $(c/a)_{int}$ ratio of Ni sublattice to 1.21 (black
rectangles in Fig. \ref{fig:fig1}(e)). Thus the basic building blocks
in Ni and Mn-Ga sublattices are not exactly tetragonal but exhibit
very small monoclinic distortion.

  Similar stabilizing mechanism as in 4O can be expected also in 10M
  and 14M structures, since they also contain nanotwin
  doublelayers. We found the difference of total energies between the
  geometrically constructed $(3\overline{2} )_{2}$ \cite{Niemann16}
  and fully relaxed 10M structure which is 2.22 meV/atom; for
  $(5\overline{2} )_{2}$ and relaxed 14M it is 1.62 meV/atom. In
  analogy with the 4O structure we suggest that the largest
  contribution to this energy reduction originates from the shifts of
  Mn and Ga at the nanotwin boundaries. For fully relaxed 10M
  structure, these shifts are indicated by red arrows in
  Fig. \ref{fig:fig2}(f).

  The alternating shifts of atomic positions in Mn-Ga sublattice
  result in significant reduction of the total energy and thus are
  crucial for stabilization of all structures containing nanotwin
  doublelayer (4O, 10M and 14M). The shifts of atomic positions also
  explain why harmonic description of modulation is insufficient for
  full description of the structure. As consequence of these shifts
  the Ni and Mn-Ga sublattices in 4O structure exhibit different
  internal $(c/a)_{int}$ ratio of their corresponding tetragonal
  building blocks (Fig. 1e). Detailed analysis of interatomic
  distances indicates different interactions between Ni atoms and
  Mn/Ga atoms and formation of very strong Ni-Mn and Ni-Ga bonds in
  the nanotwin boundary plane.

Due to higher symmetry the alternating shifts cannot be realized in NM
structure since each atom has the same number of nearest neighbours
and the strong bonds cannot be formed. In 10M or 14M structures the
shifts and strong bonds can be formed but only in nanotwin boundaries
enclosing nanotwin doublelayer, Fig. \ref{fig:fig1}(b) and (c)). Only
the 4O structure has enough flexibility to allow this distortion in
whole lattice which significantly reduce the total energy. For
stress-induced 14M$\rightarrow$NM transformation the direct growth and
annihilation of fully formed doublelayers was confirmed by in-situ
observation in TEM \cite{Ge15}.

The energy of modulated 10M and 14M structures, consisting of doublelayers
interleaved with NM structure, is similar to the energy of the
pure NM structure, Fig.~\ref{fig:fig2}(e). This seems to be in
contradiction with our statement on the beneficial contribution of
doublelayers and associated lattice distortion to the total energy. To
get the right interpretation of the observation, one need to consider
that the energy of particular structure is the result of competition
between positive contribution of nanotwin boundary energy
\cite{Gruner16} and negative contribution of boundary-boundary
interaction across the doublelayer. In 10M and 14M, the two boundaries
enclosing the doublelayer can interact only with each other across the
short distance of two atomic planes. In 4O, the number of mutual
boundary-boundary interactions per doublelayer is two times larger
than that in 10M and 14M.  From this reasoning we conclude that the
boundary-boundary interaction energy must be negative, and it is
immediately seen why the 4O structure is a distinct case with much
lower energy than 10M or 14M. However, also other factors as structure
symmetry or magnetic interactions have to be taken in the account for
full understanding of energy differences between martensitic
structures.

\begin{figure}
\includegraphics[width=86mm]{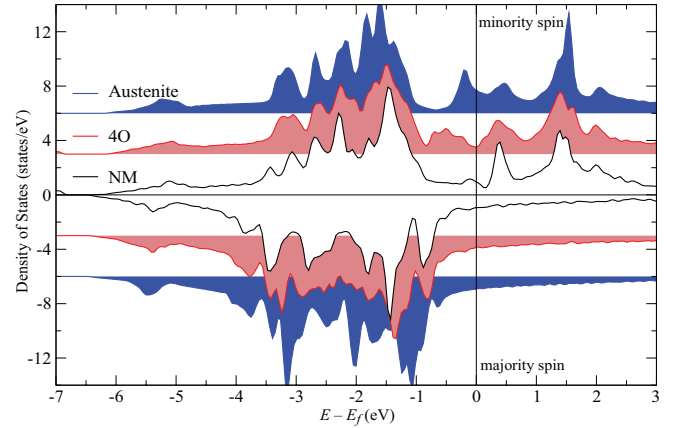}
\caption{\label{fig:fig3}(Color online)  Density of states (DOS) for NM, 4O and 
austenitic structure. The zero energy corresponds to the Fermi level $E_{f}$.  
}
\end{figure}

The clear evidence for the 4O stability is provided by the calculated
DOS displayed in Fig. \ref{fig:fig3}.  In contrast with DOS of
austenite or NM martensite, the Fermi level, $E_{f}$, lies exactly in
the middle of a pseudo-gap for minority spin channel, separating
bonding and antibonding states. The austenite exhibits the Jahn-Teller
peak right below the $E_{f}$ (blue curves in Fig. \ref{fig:fig3}),
which splits due to the tetragonal distortion of NM
structure. However, the states at $E_{f}$ still remain occupied in NM
(black curves in Fig. \ref{fig:fig3}). Similar partial occupation at
$E_{f}$ has been reported also for 10M and 14M structures (e.g. Fig. 8
in Ref.  \cite{Niemann12}).  The lowest occupation at $E_{f}$ arises
uniquely in 4O structure (red curves in Fig. \ref{fig:fig3}).  As
discussed above, it originates from the strong bond in
nanotwin-boundary plane, significantly lowering the total energy.

To get deeper insight into the stability of the 4O structure, we
employed Generalized Solid State Nudged Elastic Band (G-SSNEB) method
\cite{ssneb} to find transformation path from austenite to 4O
structure. This method allows determining the pathways of solid-solid
transformations and energy barriers involving both atomic and
unit-cell degrees of freedom with the knowledge of starting and final
lattice only. In our case the structures between austenite and 4O
martensite were linearly interpolated by ten images and then each
image was relaxed with respect to pathway coordinate by G-SSNEB
procedure to obtain minimum energy path. We have found a very tiny
energy barrier (in order of calculation error) between austenite and
4O. Using the same G-SSNEB procedure to calculate transformation
between austenite and NM martensite we obtained the barrier and
transformation path corresponding to the tetragonal deformation path
already described in literature \cite{Zayak03, Entel06, Zayak05,
  Zeleny15} confirming the validity of this approach. The barrierless
transformation between austenite and 4O is not surprising as
austenite, unstable at 0 K, transforms easily to low-symmetry ground
state due to the Fermi surface nesting \cite{Zayak03, Entel06,
  Bungaro03, Zayak06} and Jahn-Teller effect \cite{Fuji89, Brown99,
  Kart08}.

Question arises why the 4O structure has not yet been reported
experimentally in Ni-Mn-Ga. First of all, our calculations were
limited only to stoichiometric Ni$_{2}$MnGa at 0 K. The stability of
4O will be strongly influenced by temperature, impurities and
compositional disorder as these can destabilize nanotwin
doubleayers. In the Ni-Mn-Sn alloy \cite{Zheng13}, the 4O is
destabilized by increasing Mn content, which results in 4O, 10M, 14M
and NM sequence \cite{Zheng13}. From the point of view of the theory
of adaptive martensite\cite{Kaufmann10} the 10M or 14M martensites can
be seen as accommodated 4O structure with stacking errors.

In summary, based on \textit{ab initio} calculations we propose stable
4O phase in Ni$_2$MnGa, which is formed by nanotwin doublelayer
stack. The lowest total energy of the 4O phase indicates that the
nanotwin doublelayer is the most stable building element of Ni-Mn-Ga
modulated martensites and the nanotwinning or modulation is natural
property of low-temperature Ni$_{2}$MnGa phases. The 4O structure is
mainly stabilized by the shifting of Mn and Ga atoms at the nanotwin
boundary due to the different types of interactions in Ni and Mn-Ga
sublattices.  We found that similar shifting exist also in nanotwin
boundaries of 10M and 14M structures.  Thus the presence of nanotwin
doublelayers in modulated martensites reduces considerably the total
energy. This explains why stable modulated phases always include the
nanotwin doublelayers.  Although the 4O structure in Ni-Mn-Ga has not
yet been confirmed experimentally, it brings alternative view of
martensitic structures which can be relevant even for other
nanotwinned materials.

\begin{acknowledgments}

This research has been supported by the Ministry of Education, Youth and Sports 
within the support
programme „National Sustainability Programme I“ (Project NETME CENTRE PLUS – 
LO1202) and from the Large Infrastructures for Research, Experimental 
Development and Innovations project „IT4Innovations National Supercomputing 
Center – LM2015070“ and by the Czech Science Foundation (Projects No. 
14-22490S and No. 16-00043S).  
\end{acknowledgments}


\end{document}